\documentclass{tcibook}
\usepackage{fancyhea}
\usepackage{work}
\usepackage{lineno}
\usepackage{bm}       
\usepackage{graphicx}
\usepackage{hyperref}      
\usepackage{color,xspace,cite}

\newcommand{\nc}{\newcommand}  

\def\beq{\begin{equation}}
\def\eeq#1{\label{#1}\end{equation}}
\def\eeqn{\end{equation}}

\newenvironment{Eqnarray}%
   {\arraycolsep 0.14em\begin{eqnarray}}{\end{eqnarray}}
\def\beqa{\begin{Eqnarray}}
\def\eeqa#1{\label{#1}\end{Eqnarray}}
\def\eeqan{\end{Eqnarray}}

\nc{\ra}{\rightarrow}  
\nc{\slsh}{\slash\hspace*{-0.22cm}}
\def\Re{{\cal R \mskip-4mu \lower.1ex \hbox{\it e}\,}}
\def\Im{{\cal I \mskip-5mu \lower.1ex \hbox{\it m}\,}}

\nc{\vev}[1]{ \left\langle {#1} \right\rangle }
\nc{\bra}[1]{ \langle {#1} | }
\nc{\ket}[1]{ | {#1} \rangle }
\nc{\fb}{\,{\rm fb}^{-1}}
\nc{\ev}{{\rm eV}}
\nc{\kev}{{\rm keV}}
\nc{\Mev}{{\rm MeV}}
\nc{\gev}{{\rm GeV}}
\nc{\tev}{{\rm TeV}}
\nc{\mev}{{\rm MeV}}

\def\del{\partial}
\def\Dslash{\not{\hbox{\kern-4pt $D$}}}
\def\dslash{\not{\hbox{\kern-2pt $\del$}}}
\def\pslash{\not{\hbox{\kern-2pt $p$}}}
\def\ETmiss{ \not{\hbox{\kern-4pt $E$}}_T }

\def\msb{{\bar{\ssstyle M \kern -1pt S}}}

\begin{document}

\def\bibname{References}
\bibliographystyle{plain}

\raggedbottom

\pagenumbering{roman}

\parindent=0pt
\parskip=8pt
\setlength{\evensidemargin}{0pt}
\setlength{\oddsidemargin}{0pt}
\setlength{\marginparsep}{0.0in}
\setlength{\marginparwidth}{0.0in}
\marginparpush=0pt



\renewcommand{\chapname}{chap:intro_}
\renewcommand{\chapterdir}{.}
\renewcommand{\arraystretch}{1.25}
\addtolength{\arraycolsep}{-3pt}

\newcommand{\trademark}{\textsuperscript{\texttrademark}\xspace}

\thispagestyle{empty}
\begin{centering}
\vfill

{\Huge\bf Planning the Future of U.S. Particle Physics}

{\Large \bf Report of the 2013 Community Summer Study}

\vfill

{\Huge \bf Chapter 9: Computing}

\vspace*{2.0cm}
{\Large \bf Conveners: L.~A.~T. Bauerdick and S. Gottlieb}
\pagenumbering{roman}

\vfill

{\large  Study Conveners: M. Bardeen, W. Barletta, L.~A.~T.~Bauerdick, R. Brock,
D.~Cronin-Hennessy, M.~Demarteau, M.~Dine, J.~L. Feng, M. Gilchriese,
S. Gottlieb, J.~L.~Hewett, R. Lipton, H.~Nicholson, M.~E. Peskin,
S. Ritz, I.~Shipsey, H. Weerts}\\
\vspace{1cm}

{\large Division of Particles and Fields Officers in 2013:
J.~L. Rosner (chair), 
I. Shipsey (chair-elect), 
N. Hadley (vice-chair),
P. Ramond (past chair)}\\
\vspace{1cm}

{\large Editorial Committee:
R.~H. Bernstein,
N. Graf,
P. McBride,
M.~E. Peskin,
J.~L. Rosner,
N.~Varelas,
K. Yurkewicz}

\vfill

\end{centering}

\newpage
\mbox{\null}

\vspace{3.0cm}

{\Large \bf Authors of Chapter 9:}

\vspace{2.0cm}
 {\bf L.~A.~T. Bauerdick, S.~Gottlieb},
G. Bell,
K. Bloom,
T. Blum,
D. Brown,
M. Butler,
A. Connolly,
E.~Cormier,
P. Elmer,
M.~Ernst,
I.~Fisk,
G.~Fuller,
R. Gerber,
S. Habib,
M. Hildreth,
S. Hoeche,
D. Holmgren,
C. Joshi,
A. Mezzacappa,
R. Mount,
R. Pordes,
B. Rebel,
L. Reina,
M.~C.~Sanchez,
J. Shank,
P. Spentzouris,
A.~Szalay,
R.~Van~de~Water,
M.~Wobisch,
S.~Wolbers

 \tableofcontents

\newpage

\pagenumbering{arabic}

%
 
\setcounter{chapter}{8}
\chapter{Computing} 

\begin{center}\begin{boldmath}



\begin{center}

\begin{large} {\bf Conveners: L.~A.~T. Bauerdick and S.~Gottlieb}\end{large}

G. Bell,
K. Bloom,
T. Blum,
D. Brown,
M. Butler,
A. Connolly,
E. Cormier,
P. Elmer,
M.~Ernst,
I.~Fisk,
G.~Fuller,
R. Gerber,
S. Habib,
M. Hildreth,
S. Hoeche,
D. Holmgren,
C. Joshi,
A. Mezzacappa,
R. Mount,
R. Pordes,
B. Rebel,
L. Reina,
M.~C.~Sanchez,
J. Shank,
P. Spentzouris,
A. Szalay,
R.~Van~de~Water,
M.~Wobisch,
S.~Wolbers


\end{center}



\end{boldmath}\end{center}


\section{Introduction}

Computing has become a major component of all particle physics experiments
and many areas of theoretical particle physics. 
Progress in particle physics experiment and theory will require significantly more
computing, software development, storage, and networking, with different
projects stretching future capabilities in different ways.  
As a result of considerable work throughout the Snowmass process, we
recommend improved training, more community planning, careful and continuing 
review of the topics outlined in more detail below, and expanded efforts 
to share our expertise with and learn from experts beyond our field.

The Computing 
subgroups covered user needs and infrastructure.
Experimental user needs subgroups included those for
the Cosmic, Energy and Intensity Frontiers.
Theory subgroups covered accelerator science,
astrophysics and cosmology, lattice field theory, and perturbative QCD.
Four infrastructure groups predicted computing trends and
how they will affect future costs and
capabilities. These groups focused on distributed computing and facility
infrastructures; networking; software development, personnel, and training; and
data management and storage. 
The Computing
subgroups engaged with the other frontiers to learn of
their plans and to estimate their computing needs.  The infrastructure groups
engaged with vendors, computer users, providers, and technical experts. 

Our study group
considered the hardware and software needs of particle physics for the next
ten to twenty years, and recommends, not a grand project or two, but a continuing
process of monitoring and supporting the hardware and software needs of the field
in order to optimize the contribution of computing to scientific output.
One difference between computing and other enabling technologies is that there is
a vibrant global computer industry that is constantly creating new products
and improving current ones.  Although some of the computing equipment that we deploy
is customized, such as application-specific integrated circuits, the vast
majority of it is widely available.  We are not building a bespoke detector with
a multi-decade lifetime.  For the most part, we are purchasing commercially available
computing equipment that will last less than ten years.  However, we need to
carefully examine our computing needs and the available technology
to ensure that our systems are configured to cost-effectively meet those needs.
In contrast to the short lifetime of the hardware, the
software that is developed for both experiment and theory has a longer lifetime. 
However, the software is undergoing continual development and optimization.   

Two different styles of computing are currently used in particle physics.
The experimental program mainly relies on distributed
high-throughput computing (HTC). 
The distributed computing model was pioneered by Energy Frontier
experiments.  It relies on distributed computing
centers that are part of the Open Science Grid in the U.S.,
with additional centers across the globe. 
The theoretical computing and simulation needs are more commonly
addressed by high-performance computing (HPC) in which many
tightly coupled central processing units (CPUs) are working together on a single
problem. These resources are provided mostly through DOE and NSF
supercomputing centers.

One important issue to consider is 
to what degree can or should data-intensive applications that have traditionally 
relied on HTC use national supercomputer centers, which have traditionally 
been designed for HPC usage?
Work is proceeding to determine how well and how cost-effectively these HTC
applications can run at HPC centers.
Also, traditional HPC applications are developing more data-intensive
science needs, which are currently not a good match to existing
and next-generation HPC architectures. Computational resources will have to
address the demands for greatly increasing data rates, and the increased
needs for data-intensive computing tasks.

Another pressing issue facing both HTC and HPC communities is that
processor speeds are no longer increasing, as they were for
at least two decades. Instead, new chips provide multiple
cores. Thus, we cannot rely on new hardware to run serial codes faster.
We must therefore parallelize codes to increase application performance. 
Today's computer servers contain one or more multi-core chips.  Currently,
these chips  have up to 10 (Intel) or 16 (AMD) cores. 
For additional performance the server may contain
computational accelerators such as graphical processing
units (GPUs) or many-core chips such as the Intel Xeon Phi that can have up
to 61 cores. 
In the past, computing resource needs for Energy Frontier experiments scaled
roughly with the rate that processor speeds increased, following Moore's
law. In the future, this requires full use of multiple-core and many-thread 
architectures. Also,
scaling of disk capacity and throughput is of
significant concern, as per-unit capacities will no longer increase as rapidly
as they have in the past.

These changes in chip technology and high-performance system architectures
require us to develop parallel algorithms and codes, and to train personnel
to develop, support and maintain them. Different subgroups are at different
stages in their efforts to port to these new technologies. 
In the U.S., the effort to write lattice QCD codes, for
example, started in 2008 and there has been code in
production for some time; however, there
are other parts of the code that are still only running on CPUs.
Cosmological simulations have exploited GPUs since 2009, and some 
codes have fully incorporated GPUs in their production versions, running 
at full scale on hybrid supercomputers.
Accelerator science is also actively writing codes for GPUs. Some of the
solvers and particle-in-cell infrastructures have been ported and very
significant speed-ups have been obtained. The perturbative QCD community
has also started using GPUs.

These trends lead to vastly increasing code and system complexities. For
example, only a limited number of people in the field can program GPUs. In
this and other highly technical areas, developing and keeping expertise in
new software technologies is a challenge, because well-trained personnel and
key developers are leaving to take attractive positions in industry.
Continued training is important.  There are training materials
from some of the national supercomputing centers.  Summer schools are
organized by the Virtual School of Computational Science and
Engineering (www.vscse.org) and other groups. 
We must examine whether these provide the
right training for our field and whether the delivery mechanisms are
timely.  On-line media, workbooks and wikis were suggested to enhance
training. Another area of common concern is the career path of those who
become experts in software development and computing. We should help
young scientists learn computing and software skills that are marketable
for non-academic jobs, but it is also important that there be career paths
within particle physics, including tenure-track jobs, for those working at
the forefront of computation.

Subsequent sections of this chapter summarize the finding of each of our subgroups.
We start with the needs of the Energy, Intensity, and Cosmic Frontiers.  We then
turn to the needs of theoretical areas: accelerator science, lattice field theory,
and perturbative QCD.  Theoretical work in astrophysics and cosmology is included
in the Cosmic Frontier section.  
Our last section briefly summarizes our conclusions.
Additional details appear in individual sections and in the full subgroup
reports.

\section{Computing for the Energy Frontier}

Computing for experiments at the Energy Frontier is now dominated by
the huge data processing and analysis demands for the Large Hadron Collider
(LHC). The scale of the LHC computing problem has required the creation of the
global computing infrastructure of the 
worldwide LHC Computing Grid (WLCG), which has been
hugely successful.  
In each of the LHC experiments, 
data samples at the 100 petabyte scale must be accessed by experimenters around 
the world and shared between different science groups within the overall 
data analysis organization.  This has been accomplished by developments 
in networking based on a tiered system of computing centers at various 
scales. Both data storage and processing are distributed through 
this hierarchy of centers to maximize usability and throughput.

\subsection{Current and future computing needs}
Progress in distributed HTC, high-performance networks, distributed data 
management, remote data access, and work flow systems has helped 
experimental groups, production teams and scientists worldwide to
effectively access 
and use their data. Collaboration, facilitated by groups such as 
the WLCG and national consortia, enables this progress.
In the U.S.  the Open Science Grid is  bringing together the sites, 
experiments, infrastructure
providers, and computing specialists, that are necessary sustain and
further develop this distributed environment.

LHC computing today routinely uses 250,000 CPU processor cores and nearly 170
PB of disk storage in addition to large multi-hundred PB capacity
tape libraries.  The experiments generate over 1 PB per second of data
at the detector device level. Triggering and real-time event filtering  is
used to reduce this by six orders of magnitude. 
The upcoming run of LHC experiments will have a final rate to persistent
storage of around one gigabyte per second. 
The main requirement limiting the rate to storage is that of keeping 
the storage cost, and the  cost of the computing to analyze
the stored data, at a tolerable level.

Looking forward, the increased luminosity at the
HL-LHC stands out as a significant challenge. 
The expected increases in trigger rate, pileup
and detector complexity (number of channels) could increase the data rates by
a about a factor of 10 or more.   This order of magnitude increase in storage
and CPU requirements presents a new challenge for the computing infrastructure,
and the community will need time to prepare for it. The LHC community is
beginning to review their computing models as they make plans for the next
decade.  It is anticipated that the general design will be an evolution from
the current models, with the computing resources distributed at computing
centers around the world.
In contrast to the LHC and its future upgrade, science at Energy Frontier 
lepton colliders is unlikely to be constrained by computing issues.  

The full report on Computing for the Energy Frontier \cite{Fisk:2014lia}
presents a  prediction 
of the magnitude of changes that should be expected over the coming decade. 
It reviews  the changes between the Tevatron and LHC over the past 10 years. 
We argue
that the resources needed for LHC Run2, starting in 2015 and ending
in 2021, can probably be
accommodated with a roughly flat budget profile. However, the start of HL-LHC
will be a large disruptive step, like the one going from the Tevatron to the
LHC.

\begin{figure}[htb]
\begin{center}
\includegraphics[width=0.5\hsize]{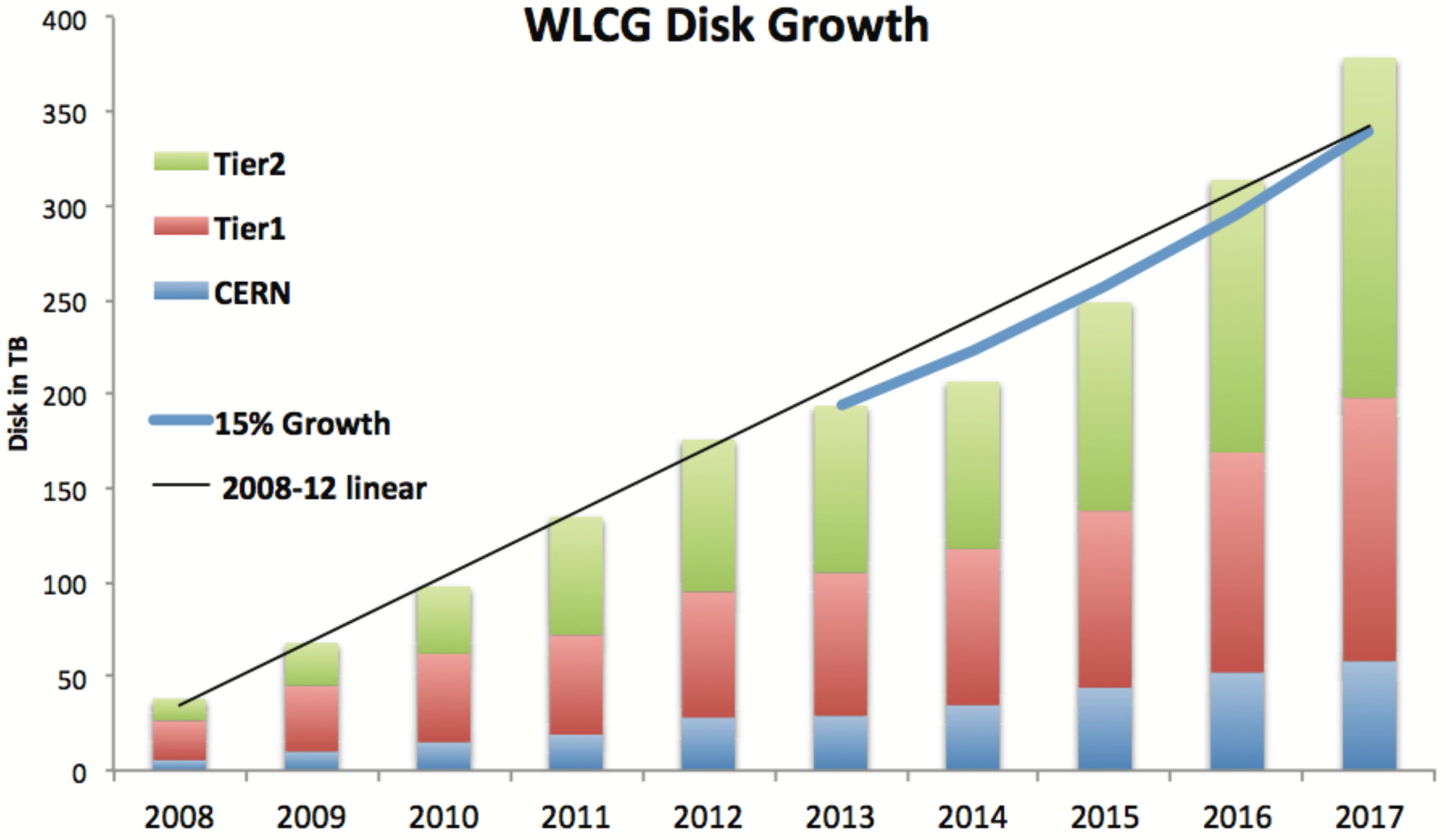}\ \ \ \
\includegraphics[width=0.45\hsize]{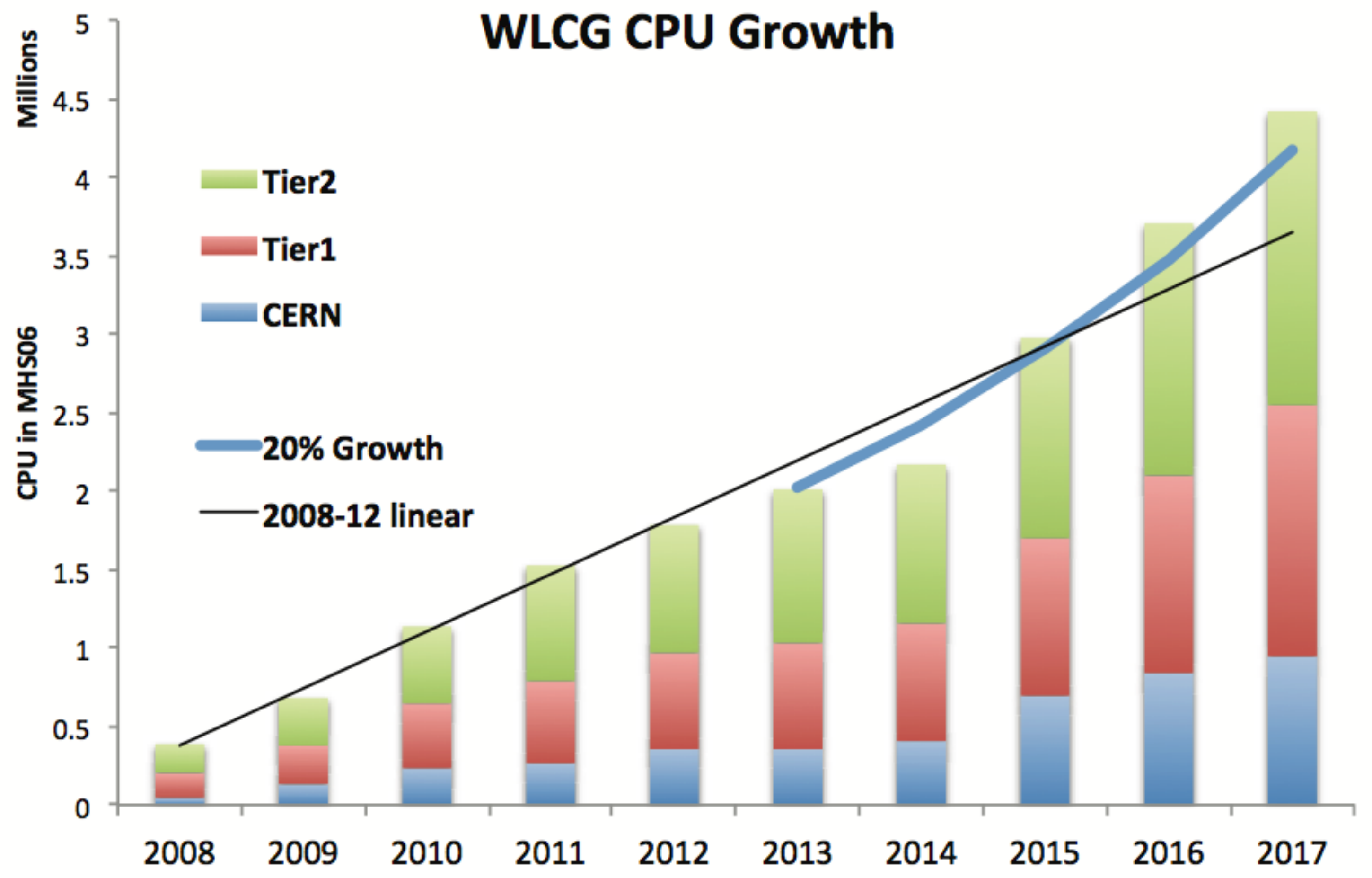}
\caption{The CPU and disk growth through the first 6 years of the LHC program
and projections to 2017.}
\label{fig:growth}
\end{center}
\end{figure}

The increases in LHC computing and disk storage since its start are shown in
Figure~\ref{fig:growth}.  CPU performance is measured in terms of a standard
benchmark known as HEP-SPEC06 (HS06) \cite{HS06}.
The CPU increases at a rate of 363K HS06 per year and
the disk at 34 PB a year on average.  The rough linear increase in CPU/yr
is the combination of three separate periods that average to linear.  The period
2008 through 2010 covered the procurement ramp for LHC as the scale of the
available system was tested and commissioned. The period from 2010 to 2013 covered
the first run, 
where the computing and storage needs increased at a rate defined by 
the volume of incoming data to be processed and analyzed.

The resources needed to accommodate
the higher trigger rate and event complexity expected in the second run define
the requirements for
2015.  The three periods roughly average out to a linear increase in
CPU power and disk capacity.

The growth curves below do not scale with total integrated luminosity but
indicate that more hardware is needed per unit time as trigger rates and event
complexity increase. It is not reasonable to expect that the techniques
currently used to analyze data in the Energy Frontier will continue to scale indefinitely.
The Energy Frontier will need to adopt new techniques and methods moving forward.

Extrapolating the growth trend out 10 years,  LHC computing would have
roughly three times the computing expected in 2015, which is lower than that
predicted by Moore's law.  LHC would reach nearly 800 PB of disk 
space by
2023, which again is roughly a factor of three greater that predicted for
2015.   These increases could
probably be achieved with close to flat budgets.   There are potential
efficiency improvements and new techniques that will be discussed below.

The luminosity and complexity increase dramatically
going to the HL-LHC or other proposed hadron collider programs.
If those plans are realized,  computing would not be on
the curve in Figure~\ref{fig:growth}, but would require a significant shift.
To estimate the increase in resources needed to move from the LHC to the
HL-LHC, it is instructive to examine the transition from the Tevatron to the
LHC.  Data rates are about a factor of ten larger for the LHC at the end
of 2013 than for the Tevatron in 2003.
However,  during that time the total computing
capacity went up by a factor of thirty, while the disk capacity, the local data
served, the wide area networking from the host lab, and the inter-site
transfers all increased by a factor of 100 to accommodate this step.  The step
from LHC Run2 to the HL-LHC will similarly require very significant additional
resources, and quite possibly a disruptive change in technologies and
approaches.
We identify two trends that will potentially help with this: the
increased use of specialized hardware, and providing and using computing as a
service.

\subsection{Trends to specialized systems and computing as a service}
The Energy Frontier will need to evolve to use alternative computing
architectures and platforms such as GPUs and other co-processors,  low-power
``small'' cores, etc., as the focus of industry development is moving away from
the classic server CPU.  Using GPUs introduces significant diversity to the
system, complicates the programming, and changes the approaches used in
scientific calculations, but can increase performance by orders of magnitude
for specific types of calculations.  Co-processors have similar potential
improvement gains, but also increase the diversity and complexity of the
system, and pose additional programming challenges.  Low-power mobile
platforms are most interesting when they are combined into massively parallel,
specialized system in which a single rack 
may have the same number of cores as a
remote computing center does today.  These systems would be used more like a
supercomputer and less like a batch farm, which will require new
expertise in dealing with these more highly interconnected computers.

Specialized hardware and architectures are likely to be deployed initially in
extremely well controlled environments, like trigger farms and other dedicated
centers, where the hardware can be controlled and specified. The next phase is
likely to be schedulable, dedicated, specialized systems that permit large-scale
calculations to achieve a goal similar to making a supercomputer center
request.  Large-scale clusters of specialized hardware owned by the experiment
are likely to come last, and are only likely to come if they can completely
replace a class of computing resources and perform a function at a reduced
cost and higher efficiency.

The other trend impacting Energy Frontier computing is the move to computing as a service
and other ``cloud-based'' solutions.  Currently,  commercial offerings, academic
resources, and opportunistic resources are all being offered through cloud
provisioning techniques.  Opportunistic resources are computers with some amount
of idle unused capacity that can be accessed by communities for free
except for the effort required to make them useful.
While commercial solutions are still more expensive
than well-used dedicated resources, there is a steady decrease in the pricing.

Energy Frontier computing should expect a transition to more shared and opportunistic
resources provided through a variety of interfaces, including cloud
interfaces.   Effort is needed to allow the community to make effective use of
the diverse environments and to perform resource provisioning across
dedicated, specialized, contributed, opportunistic, and purchased resources.

\subsection{Becoming more selective}
There are considerable concerns regarding the enormous increase in data
produced by the next round of Energy Frontier colliders,   to be processed by the offline
computing systems.  We observe that while the Energy Frontier processing capacity has
increased largely as would be expected from Moore's law and relatively
flat budgets, the storage requirements have grown much faster.  The larger
number of sites and the need for local caches, the increase in trigger rates,
and the larger event sizes drive the need for disk-based storage.

For Energy Frontier discovery physics and searches there is a case for storing all
potentially interesting events, and then computationally  applying various
hypotheses to look for new physics.  
Some searches, and many measurements, may benefit from a
new approach where much more of the processing and analysis is done with the
initial data collection and only synthesized output is archived.
This approach has the
potential for preserving physics while reducing the offline processing and
storage needs.  

We expect a change of mentality, moving away from the approach that higher
trigger rates are always better, and that all data from  all triggered events
need to be kept.   As the Energy Frontier trigger rates go up by an order of magnitude,  as
expected for the LHC, and certainly for the HL-LHC,  the experiments
should expect to be more selective in what classes of events will be fully
reconstructed,  and instead develop an approach of on-demand reconstruction,
calibration, and analysis.

Simulation and raw-data reconstruction produce derived data that can
entirely be reproduced. At the LHC, already many of the intermediate steps are
treated as transient data.   More of the data analysis steps should be moved
into the production chain, only storing the final output, with the
understanding that it can be re-derived if required later.

\subsection{Data management}
With the expected increased diversity of computing resources, Energy Frontier computing
needs to develop a data management system that can deal with all kinds of
resources.   In the next decade computing processing for the Energy Frontier will evolve to
be less deterministic, with more emphasis on cloud-provisioned resources,
opportunistic computing, local computing, and volunteer computing.   A data
management system is needed to handle the placement of the data and allow the
operations team and analysis users to concentrate more on execution of work
flows and less on placement and location of data.

Industry has put a focus on delivering content either through Content
Delivery Networks (CDNs) or through peer-to-peer systems.  The experiment data
management systems need to evolve to be more flexible in terms of what 
computing resources can be used to solve each type of computing problem,
in order to make efficient use of the diverse landscape of resoucres to which
the experiment computing modes will have to adapt.
The development of a data intensive content delivery network should not be
unique to one experiment, and should even be applicable to several scientific
domains, but this will require commitment and effort to develop.

Additional details may be found in the full subgroup report 
\cite{Fisk:2014lia}.

\section{Computing for the Intensity Frontier}

Computing at the Intensity Frontier has many significant challenges. The
experiments, projects, and theory all require demanding computing capabilities
and technologies.  Though not as data-intensive as the LHC experiments, the Intensity Frontier
experiments have significant computing requirements for simulation,  theory
and modeling, beam line and experiment design, triggers and DAQ, online
monitoring, event reconstruction and processing, and physics analysis.  It is
critical for the success of the field that Intensity Frontier computing be up-to-date, with adequate 
capacity and support, and able to take advantage of the latest
developments in computing hardware and software.

\subsection{Scope}
The Intensity Frontier encompasses a large spectrum of physics, including quark flavor
physics,  charged lepton processes, neutrinos, baryon number violation,  new
light weakly-coupled particles, and nucleons, nuclei, and atoms.  The
requirements and resources of quark flavor physics, as in Belle II and LHCb,
are similar to those of the Energy Frontier. 
Intensity Frontier experiments are carried out at a number of laboratories 
around the world, including JLab, IHEP, KEK, J-PARC, and PSI.  
Our group looked most intensely at the complex of Intensity Frontier 
experiments planned for Fermilab and the issues in finding common 
computing solutions for these experiments.  
We hope that the insights we present here will also be relevant in a 
broader, global, context.

The Intensity Frontier has increasingly become a focus of the U.S.-based particle physics program. Many  of
the experiments are designed to use very intense particle beams to measure
rare processes. There is a large number of experiments, a range of scales
in data output and throughput, and a range in the number of experimenters.
This situation can potentially lead to fragmentation, duplication of effort, lack of
access to computing advances, and higher cost than necessary to support these experiments. 
Furthermore there is significant overlap of human resources among experiments, making any significant
divergence of  software, frameworks and tools between them particularly
inefficient.  A broad range of experiments leads to a broad range of needs in 
terms of number of experimenters and the sizes of the data sets. Experiments'
computing requirements range from high-intensity, real-time processing with 
small data sets to stored large data sets that are themselves equivalent in 
size to the previous generation of collider experiments. 

Over the
last few years there has been a significant effort by the Intensity Frontier experiments at
Fermilab to join forces in using a more homogeneous set of software packages,
frameworks and tools to access infrastructure resources. This trend has
reduced fragmentation and led to more efficient use of resources. We recommend
expanding this trend to the broader Intensity Frontier community, adapted to the needs of each
collaboration.

\subsection{Survey}
For this report a qualitative survey was conducted of the current and near-term 
future experiments in the Intensity Frontier in order to understand their computing needs
and also the expected evolution of these needs.  Computing liaisons and
representatives for the LBNE, MicroBooNE, MINER$\nu$A, MINOS$+$, Muon $g-2$,
NO$\nu$A, SeaQuest,  Daya Bay, IceCube, SNO+, Super-Kamiokande and T2K
collaborations all responded to the survey. This does not cover all
experiments in all areas, but we consider it a representative survey of the Intensity Frontier
field.

The responses and conclusions to the survey can be grouped into five categories
as describe in detail below.  

\subsubsection{Support for software packages}
There is significant benefit to encouraging collaborative efforts
among experiments. An example is the LArSoft common simulation,
reconstruction, and analysis toolkit, used by experiments developing
simulations and reconstructions for liquid argon time projection chambers.
LArSoft is a collaborative effort that makes better use of development and
maintenance of resources, with maintenance support by Fermilab.  In
addition to software packages for simulation and reconstruction, there are
several other computing tools that are widely used in the field that should be
maintained.  These tools provide infrastructure access for code management,
data management, grid access, electronic log books, and document management.

\subsubsection{Support for software frameworks}
Efforts for common frameworks can have a significant impact, in terms of optimizing  development and support, and in
minimizing overhead on the training of experimenters.  The Fermilab-based Intensity Frontier
experiments (Muon $g-2$, Mu2e, NO$\nu$A, ArgoNeuT, LArIAT, MicroBooNE, LBNE) have
converged on ART as a framework for job control, I/O operations, and data provenance. 
Resources for the ART framework thus address needs that
exist across the experiments, such as more accessible parallelization of each
experiment's code. In fact, the primary limitation listed by users of ART was
the inability to parallelize jobs at the level of individual algorithms.  The
ability to do so will become more critical as the numbers of channels in Intensity Frontier
experiments continue to increase and the separation of signal from background
becomes more difficult due to the rare nature of the processes being examined.
This ability will also allow Intensity Frontier experiments to take advantage of the design of
modern computers containing multiple cores. Other experiments use LHC-derived
frameworks such as Gaudi, or homegrown frameworks like MINOS+, IceTray, and
RAT. The level of support for development and maintenance of such frameworks
varies. These experiments would also benefit from more access to
parallelization and professional computing support.

The survey identified the need for making consultants available to help with
software development. All experiments indicated that they would like to
put more computing professional effort toward parallelization of code,
establishing batch submission to off-site computing, establishing best
practices for writing software, software development, and  optimizing use of
Geant4. Such expertise is in high demand within the Intensity Frontier community. Existing expertise 
at Fermilab and elsewhere could fulfill this need of the
wider Intensity Frontier community if this was promoted and properly funded.

\subsubsection{Access to dedicated and shared resources}
In general, demands for computing resources of Intensity Frontier experiments are modest compared to those of the Energy Frontier
experiments.  However, those needs are not insignificant, and all experiments
require at least 1,000 dedicated batch slots to ensure timely physics results. 
NO$\nu$A alone requires 4.8 million
CPU hours per year for its simulation needs, LBNE expects to need
several PB of storage space each year during operation, and even smaller-scale
experiments like MINER$\nu$A and MicroBooNE expect to need PB-sized storage.
The full report shows a table of current and projected CPU needs. Each
experiment has periods of peak demand that follow cycles which are strongly correlated with
major conference cycles.  It is important to take the peak demand per
experiment into account when planning for resource needs. Those peaks
typically are much larger than the planned steady state usage. To meet those
demands for turnaround during peak usage, each experiment should have access to
additional resources on which it may run opportunistically.

The survey showed that support of the Fermilab-based experiments in terms of
storage and CPU is rated as excellent.  There are still issues, mostly in efficient data 
handling and script optimization, that require additional professional support. Professional support is required to
enable seamless use of resources through grid job submission, on- or off-site.
For Fermilab-based experiments, university and other national lab resources
are used in the production of Monte Carlo files. A common protocol to access
these resources such as OSG is expected in the future.

Non-U.S. experiments with U.S. participation enjoy significantly
lower levels of support. OSG provides some support of opportunistic use of
grid resources.  Without their own domestic computing resources these
experimenters need to rely either on resources in other countries, with low
priority, or on university-based resources that are shared among a broad
pool of university users from multiple disciplines. In contrast, non-U.S.\ experimenters
from T2K run intensively and very successfully on grid resources in Europe and
Canada.  In order to be competitive with analysis of data and simulation, the U.S.\ researchers must have access to dedicated resources that can
be shared with other Intensity Frontier experiments. It was widely noted that the lack of dedicated U.S.
resources has a detrimental impact on the science.

Networking requirements are determined by the demand that data move easily between storage systems,
be accessible for data acquisition, reconstruction, simulation and analysis,
as well as be able to take advantage of distributed computing,
either as part of the grid or cloud.  Networking must not be a barrier to
making effective use of distributed computing. With Intensity Frontier experiments becoming
larger and more international, network requirements will grow.

\subsubsection{Access to data handling and storage}
Fermilab-based experiments have
their primary data copies stored at Fermilab.  The infrastructure there
handles active storage as well as archiving of data.  The SAM system designed
and maintained at Fermilab was noted as the preferred data distribution system
for these experiments. Heavy I/O for analysis of large numbers of smaller-sized
events is an issue for systems like BlueArc. Fermilab should continue to
receive support from the DOE to ensure proper archiving of data. Other
experiments indicated using grid protocols for data storage.

All respondents indicated the need for data handling systems that seamlessly
integrate distribution of files across the network from multiple locations, to
enable experiments to make optimal use of storage resources at national labs
and universities.  The need for such a system is acutely felt by experiments
that are not based at Fermilab.  One possible solution to this problem could
resemble the tiered computing structure used by the LHC experiments, with all
Intensity Frontier experiments making use of that structure.

\subsubsection{Overall computing model and its evolution}
The computing models used in various 
Intensity Frontier experiments have a lot in common, despite large differences in
the type of data being analyzed, the scale of processing, or the specific
workflows followed.   The general model is that of a traditional event-driven
analysis and Monte Carlo simulation using centralized data stores that are
distributed to independent analysis jobs running in parallel on grid computing
clusters.  Currently, there is a remarkable overlap
in the infrastructure used by experiments. For large computing facilities such
as Fermilab, it would be useful to design a set of scalable solutions
corresponding to each of these patterns, with associated toolkits that would
allow access and monitoring. Providing resources for an experiment or changing a
computing model would then correspond to adjusting the scales in the
appropriate processing units.

Computing should be made transparent to the user, such that non-experts can
perform any reasonable portion of the data handling and simulation. All
experiments would like to see computing become more distributed across sites,
but only in very large units where it can be efficiently maintained.  Users
without a home lab or large institution require equal access to dedicated
resources. We need  continuous improvements in reducing the barrier of entry
for new users, to make the systems easier to use, and to add facilities that
help prevent the users from making mistakes.

The evolution of the computing model follows several lines including taking
advantage of new computing paradigms, like clouds; different cache schemes; and
GPU and multicore processing. There is a concern that as the number of cores
in CPUs increases, RAM capacity and memory bandwidth will not keep pace,
causing the single-threaded batch processing model to be progressively less
efficient on future systems unless special care is taken to design clusters
with this use case in mind. There is currently no significant use of multi-
threading, since the main bottlenecks are Geant4 (single-threaded) and file
I/O. Geant4's multithreading addition might have a very significant impact
across the field. There is also a possibility of parallelization at the level
of the ART framework. Greater availability of multi-core/GPU hardware in grid
nodes would provide motivation for upgrading code to use it. For example
currently we can only run GPU-accelerated code on local, custom-built systems.

\subsection{Summary}
To summarize, the computing needs of the Intensity Frontier experiments should be viewed
collectively.  When combined, these experiments require the resources and
support similar to a single Energy Frontier experiment.  The support of these experiments
directly impacts the quality of results and the efficiency with which those
results can be obtained.  There is significant support already for Intensity Frontier
experiments that are based at Fermilab and the support requirements are
expected to increase as the generation of experiments currently under
construction begin to take data.  The support of Intensity Frontier experiments that are not
based at Fermilab but still have significant U.S. collaboration, such as T2K,
needs to be improved.  Specifically, there should be an investment in
infrastructure and professional support to serve these experiments.

Transparent access to data and computing hardware resources is required for Intensity Frontier
experiments.   Users must have a simple interface with which to request data
sets that then determines the stored location of those data and returns the
data quickly to the user.  Similarly, there should be a standardized grid
submission tool that determines the optimal location for running jobs without
the user having to specify those locations.

The Intensity Frontier benefits significantly from the ability to share common frameworks and
tools, such as ART, GENIE, NuSoft and LArSoft.  The support of these efforts
must be continued and increased as new experiments come on line and more users
are added to current experiments.  Similarly, the common tools used across all
frontiers, such as ROOT and Geant4, must be supported and continuously
improved. Computing professionals are in demand as support for key software
frameworks, software packages, scripting access to grid resources and data
handling. Fermilab is a natural center for Intensity Frontier support in these areas given the
existing expertise and large number of Intensity Frontier experiments already on site.

There are efforts and problems that are shared across frontiers.  Thus, significant
investments in ROOT and Geant4 optimizations, HPC for particle physics, 
transparent OSG access, and open data solutions would have a high payoff.

Additional details may be found in the full subgroup report \cite{Rebel:2013xaa}.

 
\section{Computing for the Cosmic Frontier}

The Cosmic Frontier lies at the interface between particle physics,
cosmology, and astrophysics. Experimental and observational activities
in the Cosmic Frontier cover laboratory experiments as well as
multi-band observations of the quiescent and transient sky. Direct
dark matter search experiments, laboratory tests of gravity theories,
and accelerator dark matter searches fall into the first
class. Investigations of dark energy, indirect dark matter detection,
and studies of primordial fluctuations fall into the second class;
essentially the entire range of available frequency bands is
exploited, from the radio to TeV energies. Relevant theoretical
research also casts a very wide net --- from quantum gravity to the
astrophysics of galaxy formation.

\subsection{Experimental facilities}
The size and complexity of Cosmic Frontier experiments is also
diverse, ranging from the tabletop to large cosmological surveys,
simultaneously covering a number of precision measurements and
discovery-oriented searches. A defining characteristic of the Cosmic
Frontier is a trend towards ever larger and more complex observational
campaigns, with over a thousand researchers collaborating on sky
surveys, making them roughly the size of a large Energy Frontier
experiment. Cross-correlating different survey observations can
extract more information, help to eliminate degeneracies, and reduce
systematic errors. These factors are among the major drivers for the
computational and data requirements that we consider below.

The dramatic increase in data from Cosmic Frontier experiments over
the last decade has led to fundamental breakthroughs in our knowledge
of the ``Dark Universe'' and physics at very high energies. Driven by
technological advances, current experiments generate in excess of a
petabyte  of total data per year. The growth in data will be
continued over the coming decade by large-format CCD cameras to
measure the growth of structure from weak gravitational lensing,
wide-field spectroscopic facilities to map the clustering of galaxies,
increases in the size of direct dark matter detectors, massive radio
surveys, and ground and space-based Cosmic Microwave Background (CMB)
experiments. The mass of data will exceed 100~PB; in subsequent decades
the development of radio experiments and energy resolving detectors
will result in an increase in data streaming rates to greater than 15 GB/s.

\subsection{Simulations}
The intrinsically observational nature of much of Cosmic Frontier
science implies a great reliance on simulation and modeling. Not only
must simulations provide robust predictions for observations, they are
also essential in planning and optimizing surveys, and in estimating
errors, especially in the nonlinear domains of structure
formation. Synthetic sky catalogs play important roles in testing and
optimizing data management and analysis. The scale of the required
simulations varies from medium-scale campaigns for determining
covariance matrices to state-of-the-art simulations of large-volume
surveys, or, at the opposite extreme, small-volume investigations of
dark matter annihilation signals from dwarf galaxies.

For optical surveys, the chain begins
with a large cosmological simulation into which galaxies and quasars
(along with their individual properties) are placed using
semi-analytic or halo-based models. A synthetic sky is then created by
adding realistic object images and colors and by including the local
solar and galactic environment. Propagation of this sky through the
atmosphere, the telescope optics, detector electronics, and the data
management and analysis systems constitutes an end-to-end simulation
of the survey. A sufficiently detailed simulation of this type can
serve a large number of purposes such as identifying possible sources
of systematic errors and investigating strategies for correcting them,
or for optimizing survey design (in area, depth, and cadence). The
effects of systematic errors on the analysis of the data can also be
investigated. Because of the very low level of statistical errors in
current and next-generation precision cosmology experiments, and the
precision with which deviations from $\Lambda$CDM are to be measured,
this is an absolutely essential task.

Facilities for carrying out the required simulations include
large-scale resources at DOE and NSF supercomputing centers, augmented
by local clusters. Data-intensive computing platforms are also needed
to deal with the enormous data streams generated by cosmological
simulations. The data throughput can easily exceed that of
observations; data storage, archiving, and analysis requirements
(often in concert with observational data) are just as demanding as
for observational data sets. Although there are significant challenges
in fully exploiting future supercomputing hardware, available
resources should satisfy performance requirements, currently at the
scale of $\sim$10~PFlops. These requirements are expected to cross into
the exascale regime after 2020. The data-related issues are more
serious and will need changes in the current large-scale computing
model. Successful implementation of the recently suggested Virtual
Data Facility (VDF) capability at computing centers would go a long
way towards addressing these issues for Cosmic Frontier simulations.

Simulation requirements are projected to increase steeply. Current
allocations are estimated to be of the order of 200M compute
hours/year, with associated storage in the few PB range, and a shared
data volume of the order of 100~TB. Data management standards and
software infrastructure vary widely across research teams. The
projected requirements for 2020 are an order of magnitude increase
in data rates (to 10-100~GB/s), a similar increase in peak
supercomputer performance (200~PFlops), and the ability to store and
analyze data sets in the 100~PB class. It is difficult to make precise
estimates for 2030, as hardware projections are hazy, but the
science requirements based on having complete data sets from missions
such as LSST, Euclid, and large radio surveys would argue for at least
another order of magnitude increase across the board.

\subsection{Computational resources and architectures}
Today's architectures for data analysis and simulations include supercomputers, 
that are suitable for massively parallel computations where the number of 
cycles per byte of data is huge.  These possess a large distributed memory 
but a relatively small amount of on-line storage.
Database servers occupy the opposite range of the spectrum,
with a very large amount of fast storage, but not much processing
power on top of the data. For most scientific analyses, the required
architecture lies somewhere in between these two: it must have a large
sequential I/O speed to petabytes of data, and also perform very
intense parallel computations. 

The use of computational resources will need to grow to match the
associated data rates for the processing and analysis of observational
data and for simulated astrophysical and cosmological processes. Most
of the data processing pipelines use linear time algorithms, where the
amount of processing is roughly proportional to the amount of data
collected by the instruments.  Exceptions to this linear scaling arise, however,
with many of the algorithms that are applied to the accumulated data
including optimization and clustering methods whose computational requirements
grow as a quadratic function of the data or greater.

Most pipelines can be characterized by the number of cycles needed to
process a byte of data. Typical numbers in astrophysics today range
from a few thousand to 100K cycles, so that processing a canonical
100~PB data set requires 10$^{22}$ cycles, or about a billion CPU
hours. One particular characteristic of this processing is that it
will require a reasonable, but not excessive, sequential I/O rate to
data storage disks, typically less than a GB/s per processing compute
node.

Much of this processing is massively parallel, and thus will execute
very well on SIMD (Single Instruction, Multiple Data)
architectures. Emerging many-core platforms will therefore have a huge
impact on the efficiency of data processing pipelines. While these
platforms are harder to code for, pipeline codes will be based on
well-designed core libraries, where it will be cost-efficient to spend
resources to optimize their parallel execution, thus substantially
decreasing the hardware investment.

The projected data volumes for archiving of observational data are not
particularly large compared to commercial data sets (with the possible
exception of the Square Kilometer Array). Given that the eventual data
volumes will probably exceed a few exabytes, the analyses must be
co-located with the data.

The most likely high-level architecture for scientific analyses will
be a hierarchy of tiers, in some ways analogous to the LHC computing
model, where the (top) Tier~0 data is a complete capture of all raw data.
Derived and value-added data products are moved and analyzed
further at lower tiers of the hierarchy, which are not necessarily
co-located with the Tier~0 data centers.

The archives will have to be based upon intelligent services, where
heavy indexing can be used to locate and filter subsets of the
data. There is a huge growth in the diversity of such ``Big Data
Analytics'' frameworks, ranging from petascale databases to an array
of simpler solutions. Over the next five years a few clear winners
will emerge, allowing the Cosmic Frontier community to leverage the
best solutions. A high-speed, reliable, and inexpensive networking
infrastructure connecting the instruments and all the sites involved
in the archiving will be crucial to the success of the entire
enterprise.

Fast graph processing will become increasingly important to analyze
large and complex simulations and track complex spatio-temporal
connections among objects detected in multi-band time-domain
surveys. To efficiently execute algorithms that require large matrices
and graphs, it is likely that large (multiple TB) memory (RAM) will be
melded with multiprocessors to minimize communication overhead. Also,
new storage technologies with fast random access (SSD, memory bus
flash, phase change memory, non-volatile RAM) will play a crucial role
in the storage hierarchy.

\subsection{Data access and analysis}
Large-scale data sets, arising from both simulations and experiments,
present different analysis tasks requiring a variety of data access
patterns. These can be subdivided into three broad categories:
localized data processing, global data processing, and rendering
graphics.

Some of the individual data accesses will be very small and localized,
for example, interrogating the properties of individual halos or galaxies, and
recomputing their observational properties. These operations typically
return data in small blocks, require a fast random access, a high I/O
performance, and are greatly aided by good indexing. At the same time
there will be substantial computation needed on top of the small data
objects. These accesses can therefore benefit from a good underlying
database system with enhanced computational capabilities. Going beyond
the hardware requirements, this is an area where the clever use of
data structures will have an enormous impact on the system
performance, and related algorithmic techniques will be explored
extensively. The challenge here is that the small data accesses will
be executed billions of times, suggesting a parallel, sharded database
cluster with a random access capability of tens of millions of IOPS
and a sequential data speed of several hundred GB/s, with an unusually
high computing capability inside the servers themselves.

At the other end of the spectrum are analyses that need to access a
large fraction of all collected data, such as computing an FFT of a
scalar field over the entire volume, or computing correlation
functions of various orders, over different subclasses of
objects. These require very fast streaming access to data, algorithms
that can compute the necessary statistics over (possibly multiple)
streams, and multiprocessors that can handle these highly
parallelizable stream computations efficiently. Here the requirements
would be a streaming data rate in access of 500 GB/s between the data
store and the processing, and a peak processing capability of several
PFlops. These patterns map best onto traditional HPC systems, with the
caveat of the extreme data streaming requirements.

The third type of access pattern is related to rendering computer
graphics. These tasks will generate various maps and projections,
touching a lot of data, and typically generating two-dimensional
images. Such tasks include computing maps of dark matter annihilation
in trillion-particle simulations, ray-tracing to compute gravitational
lensing signatures over a large simulation, and generating ray-traced
simulated images for future telescopes. These ray-traced images are
based on simulations and detailed telescope and atmospheric models. As
many of these tasks are closely related to computer graphics, mapping
to GPU hardware will be very important, as this approach can yield
performance gains of well over an order of magnitude.

Dealing with each of these access patterns demands substantial
investments in hardware and software development. To build an
efficient streaming engine, all hardware and software bottlenecks must
be eliminated, since a single choke point can seriously degrade the
performance of the whole system. In terms of algorithms, many
traditional RAM-resident algorithms must be recast into streaming
versions. A rethink of statistical algorithm design is needed, and
computations (and computability) should be explicitly included into
the cost tradeoffs.

Table \ref{tab:CosmicCompNeeds} summarizes current and future
computational needs for the Cosmic Frontier.

\subsection{Development and support of a computational community}
The need for better programming models and better high-level
abstractions is evident. In a complex, massively parallel system it
will become increasingly difficult to write code explicitly
instructing the hardware. Therefore, there is a need to explore and
embrace new declarative programming models where the explicit
execution of the code is transparent to the user. At a higher level,
there is a pressing need for the development of a sustainable software
effort that can provide a baseline of support to multiple experiments,
with experiment-specific extensions being built on top of such a
capability. This will require a community effort to develop and
implement new algorithms, programming models, workflow tools, as well
as standards for verification, validation, and code testing. A
coherent plan for long-term support to maintain and further develop
the resulting software base will have to be put in place.

Directly analogous to building a community-supported software base for
Cosmic Frontier experiments, there is a related need for bringing
together larger collaborations in the area of simulations. The lattice
QCD community has shown what is possible in this direction by working
together in a large national collaboration. Such efforts are now
beginning within the Cosmic Frontier and will hopefully come to
fruition in the near term.

While much of the science in the Cosmic Frontier is undertaken by
small groups of physicists, the collaborations themselves have grown
to hundreds and sometimes thousands of members. Many of the techniques
utilized by these collaborations are common to multiple Cosmic
Frontier experiments. Most experiments have, however, developed their
analysis and processing software independently of other programs. This
can lead to duplication of effort, software that is tailored only to
meet a specific need, non-scalable approaches, and software that is
difficult to sustain beyond the lifetime of an individual
experiment. To make computing developments more robust, a sustainable
software initiative is highly desirable. A substantial community must actively
develop and deploy the tools created within such a program.

Additional details may be found in the full subgroup report \cite{Connolly:2013ibs}.

\begin{table}
\begin{center}
\begin{tabular}{|l|r|r|r|} 
 \hline 
{\bf Experimental Data} & 2013 & 2020 & 2030+ \\
\hline
Storage & 1 PB & 6 PB & 100--1500 PB \\
Cores & 10$^3$ & 70K & 300+K \\
CPU hours & 3x10$^6$ hrs & $2\times 10^8$ hrs & $\sim 10^9$ hrs \\
 \hline 
{\bf Simulations} &2013 & 2020 & 2030+ \\
 \hline 
Storage & 1--10 PB & 10--100 PB & $> 100 $PB -- 1EB\\
Cores & 0.1--1M & 10--100M &$> 1$G\\
CPU hours & 200M & $>$20G & $> 100$G\\
\hline
\end{tabular}
\caption{Computing requirements for Cosmic Frontier science over the next
 10--20 years.}
\label{tab:CosmicCompNeeds}
\end{center}
\end{table}

 
\section{Computing for accelerator science}

Particle accelerators are critical to scientific discovery, both nationally and worldwide. The development and optimization of accelerators are essential for advancing our understanding of the fundamental properties of matter, energy, space, and time. Modeling of accelerator components and simulation of beam dynamics are necessary for understanding and optimizing the performance of existing accelerators, for optimizing the design and cost-effectiveness of future accelerators, and for discovering and developing new acceleration techniques and technologies.  In addition, the combination of fast and sophisticated analytics with large-scale simulations will be very important to obtain the control-room feedback capabilities required by Intensity Frontier accelerators.

\subsection{Simulations} \label{subsec:accel-simu}
The requirements for high-fidelity computer simulations of accelerator systems and accelerator components are driven by the need to develop and optimize new accelerator concepts and design machines based on these concepts, and maximize the performance of existing accelerators based on existing concepts and technologies.  For  Energy Frontier applications this means supporting the development of new techniques that will increase the accelerating gradients so future machines are more compact and less costly. The options considered in our study include acceleration in plasma structures, using either laser- or beam-driven wakefields, dielectric structures driven by lasers or RF (GHz), the development of new lepton collider designs such as muon colliders and two-beam acceleration, and optimization of existing technologies such as superconducting rf cavities. For  Intensity Frontier applications, simulations are essential in developing and optimizing integrated designs in order to minimize beam losses.  Such losses are caused by instabilities generated either by beam self-interactions, or by interactions of the beam with the accelerator structures or other media present in the beam pipe.  The simulations considered in our study focused on designing mitigation techniques and determining optimal operational parameters.  Hadron colliders at the Energy Frontier have similar requirements, although self-interactions are not important, while beam-beam interactions (which are similarly computationally intensive) have to be included.  
Simulations of accelerators for both the Energy and the Intensity frontiers are computationally demanding because they often involve a wide range of time and length scales and a wide spectrum of interoperating physics components. For example, simulations of high-intensity proton drivers which are a few km long and operate using EM wavelengths of $10$-$100$ m,  with machine components of the order of $1$-$10$ m, must resolve particle bunches of the order of a few mm. Similarly, laser-plasma accelerators (LPA) of the order of $1$ m in length must resolve laser wavelength and electron bunch size of the order of $1$ $\mu$m.

Most software for accelerator science are already parallelized and scalable to 
more than ten thousand cores on high performance computers. 
Modeling physical fields using various approximations requires different 
numerical methods.  For example, electrostatic models utilize multigrid, 
adaptive mesh refinement (AMR) multigrid or spectral methods.
On the other hand, fully electrodynamic models use a variety of
finite difference and finite element methods.
Quasi-static models use spectral methods and particle-in-cell, among
other methods.
There are ongoing R\&D efforts to port these numerical models to new architectures such as GPU-based machines.

Progress in accelerator science requires efficient use of HPC.
Each simulation step requires communication among thousands to
millions of processors, so a fast interconnect is essential.
The major modeling applications from both Energy and Intensity
Frontiers are shown in Table~\ref{tab:CompNeeds}.  The estimate of needs is
based on the current performance of our codes on the Hopper supercomputer at
the National Energy Research Scientific Computing Center (NERSC). 
The Energy Frontier has much greater data storage and networking needs
than accelerator science, so we do not detail our needs in this area, with one
exception.  We assume that we can make use of the systems required by
the Energy Frontier.

Our user community (accelerator scientists operating machines or performing R\&D) and our own community (computational accelerator physicists and theorists) identified the need for programmatic coordination and support of code development and computing R\&D to create a sustainable
computational accelerator science program as an essential requirement for the future.   Porting of our
algorithms and workflows to new computing architectures
(light-weight CPU plus accelerator) and the R\&D necessary to
create and evaluate new algorithms is an important component of
such coordinated program (including close interactions with HPC
centers to utilize test-beds of new architectures). An example of
such programmatic support of code development today is the SciDAC program, although
it is desirable that in the future there is more focus on the
specific physics solutions needed to further develop our tools.  Another
common theme is the need for supporting the development of
community libraries and tools, including standardized user
interfaces, geometry and data descriptions, I/O and analysis tools.
Because our applications require true HPC capabilities,
it is important to develop generic workflow tools that perform in an HPC
environment as well as on local workstations and clusters. Also
important is the development and
integration to our toolkit of parameter optimization libraries,
that will be available across all HPC platforms.  The development
of such an environment will enable experimentalists and machine
operators to take advantage of these computational capabilities
and will be essential in training students and young researchers
to help develop the new accelerator concepts and technologies
that will move the field of particle accelerators forward.   In
addition, it is essential for such a program to support and
coordinate physics model validation and verification, ultimately
with comparisons to experimental data of well controlled
experiments in test facilities or operating accelerators.

\subsection{Feedback and control systems}
Intensity Frontier machines of the future require control room
feedback capabilities because of the beam-loss implications.  This 
capability is also important to Energy Frontier test
facilities for guiding and interpreting experiments.   The utilization of new computing technologies could make 
delivering  such a fast turnaround possible. The challenge on both the
performance of the computational tools and the availability of
computing resources becomes even  more daunting if we consider
the need to analyze the simulated data in order to extract useful
information.  The analysis of the simulated data ($\sim$ TB) has
to produce the same quantities observed by the beam diagnostic
detectors.  Note that this is a more general requirement, because
it is necessary for accurate comparisons of simulated and
observed data independently of the ability to do that in ``almost
real-time'' in the control room.  Analysis workflow
and synthetic diagnostic tools similar to those used by Energy Frontier
experiments have to be developed to properly model the detector
response and maintain and correlate the information of the
simulated physics variables to those ``smeared'' by modeling the
beam diagnostics.  Such analysis tools have to be HPC capable, to
allow for the fast turnaround necessary for control room
feedback, and they will also require new models
and algorithms.  Finally, this is probably the only application
in accelerator modeling that data transfer speed and data
availability, storage, and cataloging have similar requirements to
those of a  DAQ system for a particle physics experiment.

\subsection{Multi-physics modeling}
 Different applications have different specific
requirements for the development of new or more efficient physics
or computational models, but all of them require integrated
multi-scale, multi-physics modeling.  
Although the physics models implemented in today's simulation tools utilize sophisticated HPC infrastructure, because of the size of the computation, often ``single physics'' or ``few physics" models are included in a run. 
The different physics effects are studied separately, as if they were independent.  
This is not the case in general, affecting our ability to find optimal design and operational parameters.  
More efforts are needed to integrate multiple physical effects for more accurate simulations, with the ability to utilize massive computing resources beyond the capabilities of today. 
   In the Energy Frontier, where single components of the accelerator are simulated separately, end-to-end simulations and integration between components are needed.  For example, plasma-based accelerator simulations must be advanced from modeling current experiments at the 10 GeV and 0.1 micron emittance level to future collider concepts involving hundreds of stages at the 0.01 micron emittance level. 
This also requires integration of additional physical models such as scattering and radiation. For high-intensity circular proton machines, a large number of macro-particles ($\sim 10^9$) must be used in the simulations in order to accurately represent percent-level losses. In addition, detailed models of important components relevant to all frontier applications are missing from our simulation toolkits because of prohibitive computational cost and complexity.
(For example, target modeling, including gas dynamics, MHD, and heat loading/dissipation, must be integrated to our toolkit.) 
Deployment of such capabilities
will enable end-to-end simulations to validate designs based on
new concepts and end-to-end operational parameter optimization of
accelerators about to be commissioned.  It should be noted that
in some cases end-to-end modeling also involves integration of physics
and numerical models developed for different applications (for
example, for a plasma-based accelerator consisting of many plasma
stages, both plasma physics tools and conventional beam-dynamics
tools have to be used in the model to produce an optimal
solution).

\subsection{Design optimization}
Intensity Frontier accelerator needs are dominated by the need to
control and mitigate beam losses.  This demands both careful
design of the accelerator structures and accurate modeling of
beam-halo (and its creation mechanisms), the accelerator
geometry (apertures), and the positions and field strengths of each accelerator 
element.  This implies tracking many bunches of $\sim 10^9$
macroparticles per bunch for $\sim 10^5$ turns including
self-fields, impedance effects, and bunch-to-bunch interactions.
Finding the optimal parameters of operation will require
end-to-end  optimization runs, while developing mitigation
techniques possibly requires the implementation of new physics in the HPC
environment, to model the new components (for example, electron
lenses for space-charge compensation). 

Energy Frontier accelerator needs are dominated by the need to
develop end-to-end simulations to characterize and optimize beam
stability, emittance, and transport efficiency.  New
accelerator concepts have many specific new physics model
capability needs. It will be necessary to develop electromagnetic
plasma and beam methods capable of resolving 0.1 km-scale
propagation of 10 nm scale emittance bunches and laser drivers,
and the corresponding bunch conditioning and focusing. 
There are also needs common to the Energy and Intensity Frontiers:
for example, radiation and
scattering, which is relevant to muon collider, plasma and
gamma-gamma options, and modeling of targets.
Developing these new models
demands R\&D both on the physics and the numerical algorithms.
Because of the physics requirements imposed by some of the new
concepts considered, minimization of numerical noise is very
important in these applications.  This constraint has a direct
impact on the choice of numerical techniques for different
physics implementations.  Plasma accelerators additionally require 
computation of these effects with accurate plasma and laser dynamics,
 often requiring unique algorithms.

Additional details may be found in the full subgroup report \cite{Spentzouris:2013jla}.

\begin{table}[t]
\begin{center}
\begin{tabular}{|l|l|} 
 \hline 
 Computation (Mhours) & $15000$ \\ \hline
 Typical cores for production runs & $50000$ \\ \hline
 Maximum cores for production runs & $5$M\\ \hline
 Data read and written per run (TB) & $1000$\\ \hline
 Minimum I/O bandwidth & $100$ GB/sec\\ \hline
 Memory requirement per core & $0.2$ GB \\ \hline
Shared file-system space (on site) & $6$ PB\\ \hline
Shared file-system space (distributed, cataloged) & $60$ PB\\ \hline
\end{tabular}
\caption{Computing needs for accelerator science in 10 years.}
\label{tab:CompNeeds}
\end{center}
\end{table}

 
\section{Computing for lattice field theory}
\label{chap:LFT}

One of the foremost goals of particle physics is to test the Standard Model
and to search for indications of new physics beyond. In
many cases, interpretation of the experimental measurements requires a
quantitative understanding of the nonperturbative dynamics of the quarks and
gluons in the underlying process.  Lattice gauge theory provides the only
known method for \emph{ab initio} quantum chromodynamics (QCD) calculations
with controlled uncertainties, by casting the fundamental equations of QCD
into a form amenable to high-performance computing.  Thus, facilities for
numerical lattice gauge theory are an essential theoretical adjunct to the
experimental particle physics program.  Lattice QCD calculations now play
an essential role in the search for new physics at the Intensity Frontier.
They provide accurate results for many of the hadronic matrix elements needed
to realize the potential of present experiments probing the physics of
flavor. The methodology has been validated by comparison with a broad array of
measured quantities, several of which had not been well measured by experiment
at the time of the first precise lattice
calculations.  In the coming decades, lattice QCD will
play an expanded role in the search for new physics at both the Energy and
Intensity Frontiers.

The U.S. Lattice QCD Collaboration (USQCD), which consists of
most theoretical physicists in the country involved in the numerical studies
of QCD and beyond-the-Standard-Model theories, represents the lattice gauge
community.  Their efforts have been supported in an essential way by hardware
and software funding provided by the High Energy and Nuclear Physics Program
Offices of the Department of Energy.  The USQCD Collaboration's current
hardware project ends in FY2014, and the collaboration has applied for a
five-year project extension, ``LQCD-ext~II.''  The Collaboration's ongoing
software development and maintenance activities are supported by SciDAC-3
grants.

The report of the lattice field theory working group \cite{Blum:2013mhx}
summarizes the scientific
goals of the U.S. lattice gauge theory community, presents the current and
future computing needs and plans, and argues that continued support of the
U.S. (and worldwide) lattice-QCD effort is essential to fully capitalize on
the enormous investment in the particle physics experimental program.

\subsection{Lattice field theory scientific motivation}

Precision measurements at the Energy and Intensity Frontiers probe
quantum-mechanical loop effects, and are therefore sensitive to physics at
higher energy scales than those directly accessible at the LHC.  Contributions
from new heavy particles may be observable as deviations of the measurements
from Standard~Model expectations, provided both the experimental measurements
and theoretical predictions are sufficiently precise.  The scientific impact
of many future experimental measurements therefore hinges on reliable
Standard-Model predictions on the same time scale as the experiments and with
commensurate uncertainties.

For many quantities, the comparison between the measurements and
Standard-Model predictions are currently limited by theoretical uncertainties
from nonperturbative hadronic matrix elements or fundamental QCD parameters
that can only be computed numerically with lattice QCD.  The USQCD
Collaboration has laid out an ambitious vision for future lattice calculations
matched to the experimental priorities of the planned experimental particle
physics program over the next decade in the white papers ``Lattice QCD at the
Intensity Frontier'' and ``Lattice Gauge Theories at the Energy Frontier''
\cite{USQCD_IF_whitepaper13,USQCD_EF_whitepaper13}.  These detailed documents
present a concrete five-year plan for both the collaboration's foremost
scientific goals and the theoretical, algorithmic, and computational
strategies for achieving them.  The highest scientific priorities include
the following:

\begin{itemize}

\item Improving calculations of hadronic matrix elements involving
quark-flavor-changing transitions which are needed to interpret rare kaon
decay experiments

\item Improving calculations of the quark masses $m_c$ and $m_b$ and the
strong coupling $\alpha_s$ which contribute significant parametric
uncertainties to Higgs branching fractions

\item Calculating the nucleon axial form factor which is needed to improve
determinations of neutrino-nucleon cross sections for experiments such as
LBNE

\item Calculating the nucleon light- and strange-quark contents which are
needed to make model predictions for the $\mu \to e$ conversion rate at the
Mu2e experiment and to interpret dark-matter detection experiments in
which the dark-matter particle scatters off a nucleus

\item Calculating the hadronic light-by-light contribution to muon $g-2$, which
is needed to solidify and improve the Standard-Model prediction and interpret
the upcoming measurement as a search for new physics

\end{itemize}

Lattice field-theory calculations will also increasingly contribute to
collider experiments at the LHC 14-TeV run by providing quantitative
nonperturbative input for Higgs and other new-physics model building.

\subsection{Lattice field theory computing resources}

Substantial high-performance computing resources are needed to calculate
hadron masses and interactions with sufficient precision to test the Standard
Model against emerging experimental measurements.  Lattice gauge theory
simulations require parallel programming techniques, with the calculations
running cooperatively across hundreds to many thousands of processors or
processor cores.  The simulations must be run on hardware suitable for
massively parallel computations.  Although the simulations are
floating point intensive, on all current high-performance computing systems
throughput is limited by the rate that operands can be supplied to the
floating point execution units, either because of memory bandwidth limitations
or by the latency and bandwidth of interprocessor communications.
Interprocessor communications of data rely on message-passing
algorithms, typically implemented using an MPI~\cite{MPI} library.

At present, lattice theorists in the United States run these codes on a
variety of hardware.  The first type is commodity clusters based on Intel or AMD x86
processors and Infiniband networks, which have hundreds of nodes and
thousands of cores.  A second type is accelerated commodity clusters, similar to the the
standard clusters but with general purpose graphics processing units (GPUs) or
Intel Many Integrated Core (MIC) accelerators installed in each server; these
clusters have fewer nodes but typically hundreds of accelerators.  A third type is very-large-scale Cray supercomputers, consisting of thousands of AMD x86 processors
with a proprietary network, with the newest models also containing thousands
of GPUs.  Finally, lattice theorists use very large scale IBM BlueGene supercomputers, consisting of
hundreds of thousands of PowerPC cores interconnected on a proprietary
network.

Access to high-performance computing at both supercomputer ({\em capability})
and cluster ({\em capacity}) scales is essential for the lattice field theory
community.  A typical lattice-QCD analysis campaign involves a mix of problem
sizes.  The largest-scale computations are the generation of ensembles of
gauge fields, but at least as much integrated high-performance computing
capacity is required for the small- to large-scale parallel computations
(``analysis jobs'') to calculate different physical observables on these
ensembles.  In the U.S., the lattice community utilizes national
leadership-class supercomputing centers for the ensemble generation and for
the largest analysis jobs, as well as dedicated hardware purchased and
operated by USQCD for the much larger volume of small-to medium-scale analysis
jobs.

Table~\ref{tab:current} lists the leadership-class facility
capability and dedicated capacity resources utilized
for lattice-QCD simulations since 2010 by the USQCD
collaboration.  The capability resources are broken out showing both the ANL
and ORNL leadership class facilities; the capacity resources include all usage
on the DOE HEP- and NP-funded hardware at
Fermilab, Jefferson Lab, and BNL.  Subgroups within USQCD also use the DOE's
National Energy Research Scientific Computing Center (NERSC), centers
supported by the NSF's Extreme Science and Engineering Discovery Environment
(XSEDE) Program, and other facilities.

\begin{table}[t]
\begin{center}
\begin{tabular}{lccc}  
\hline\hline
Year & ANL LCF & ORNL LCF & Dedicated Capacity Hardware \\[-0.75mm]
& (BG/P + BG/Q core-hours) & (Cray core-hours) & (core-hours) \\[0.5mm]  \hline
2010 & 187M & 53.6M & 125M \\
2011 & 182M & 49.8M & 205M \\
2012 & 143M & 77.9M & 330M \\
2013 & 290M (allocated) & 140M (allocated) & 971M (planned) \\ \hline\hline
\end{tabular}
\end{center}
\caption{Utilized core-hours of leadership-class facility
(LCF) and dedicated capacity hardware for lattice-QCD simulations.  The
conversion factors for lattice-QCD sustained Tflop/sec-years, assuming 8000
hours per year, is 1 Tflop/sec-year = 3.0M core-hour on BlueGene/Q hardware, and 1 Tflop/sec-year = 6.53M core-hour on
BlueGene/P and Cray hardware. Only USQCD-Collaboration resources are shown.
The drop in ANL LCF utilized capacity in 2012 occurred
because fewer opportunistic core-hours (``zero-priority queues'') were
available due to increased demand by other facility users.}
\label{tab:current}
\end{table}

Because of the variety of processor types and parallel architectures,
efficient utilization of the above computing resources requires flexible and
effective software.  Since 2004, DOE grants to USQCD during the three
SciDAC~\cite{SciDAC} programs (2001--2006, 2006--2011, and 2011--2016) led to the
development of the USQCD software stack~\cite{SciDAC-software}.  This stack
includes low-level communications and I/O application program interfaces
(APIs) implemented via libraries ported to and optimized for each of the
architectures.  The stack includes linear algebra libraries with routines that
operate on single lattice sites, or across a full lattice with communications
between neighboring sites.  Lattice-QCD applications utilize the various
libraries of the software stack to run efficiently on any of the available
computing resources.  The USQCD software stack is a publicly available
resource supporting all of the main lattice gauge and fermion actions in
current use.  Further, it provides a general purpose framework that can be
extended to other quantum field theories besides QCD.

\begin{table}[t]
\begin{center}
\begin{tabular}{lccc}
\hline\hline  
Year & Leadership Class  & Dedicated Capacity Hardware  \\[-0.75mm] 
& (Tflop/sec-yrs) & (Tflop/sec-yrs) \\[0.5mm] \hline
2015 & 430 & 325 \\
2016 & 680 & 520 \\
2017 & 1080 & 800 \\
2018 & 1715 & 1275 \\ 
2019 & 2720 & 1900 \\ \hline\hline
\end{tabular}
\smallskip
\caption{Available resources for lattice-QCD simulations assumed for the
planned program of physics calculations.  The conversion factors for
lattice-QCD sustained Tflop/sec-years, assuming 8000 hours per year, is 1
Tflop/sec-year = 3.0M core-hour on BlueGene/Q hardware, and 1 Tflop/sec-year =
6.53M core-hour on BlueGene/P and Cray hardware.}
\label{tab:fiveyear}
\end{center}
\end{table}
  
The planned U.S. scientific program in lattice field theory over the next
five years assumes the continued availability to USQCD of capability resources
at the DOE leadership class facilities, as well as the availability of
dedicated capacity resources at Fermilab, Jefferson Lab, and BNL, deployed and
operated under the proposed ``LQCD-ext~II'' project extension.  Table~\ref{tab:fiveyear}
shows the anticipated sustained LQCD Tflop/sec-yrs provided by these resources
by year.  Completion of the planned physics calculations will require well
over an order of magnitude of increased computing capacity beyond that used in
prior years.  Use of leadership-class facilities alone would provide
insufficient computational resources and would be unsuitable for the full mix
of lattice-field-theory job requirements.  Cluster-class parallel computing
hardware, including systems with GPU accelerators, delivers capacity with the
highest cost effectiveness for jobs ranging in size from tens to thousands of
cores.

Over an order of magnitude increase in storage utilization (disk and tape)
from the current approximately 2 petabyte usage will also be needed to support
the planned simulations.  Further, the anticipated evolution of high-performance computing hardware will require the evolution of software and the
introduction and refinement of new techniques and algorithms.  Positions for
postdocs and scientific staff to develop new lattice-gauge-theory code cannot
be supported by grants to lab and university theory groups alone, but must be
augmented through grant programs such as SciDAC.

\subsection{Lattice field theory summary}

Numerical lattice-QCD calculations are needed to interpret many upcoming
experimental measurements at the Energy and Intensity Frontiers as tests of
the Standard Model and new-physics searches.  Nonperturbative hadronic matrix
elements and fundamental QCD parameters enter the Standard~Model predictions
for many processes as diverse as rare kaon decays, Higgs branching fractions,
and the muon anomalous magnetic moment.  Thus, facilities for numerical
lattice gauge theory are an essential theoretical complement to the
experimental particle physics program.

The successful accomplishment of USQCD's scientific goals requires
access to both capacity and capability machines, and hence support for both
leadership-class facilities and dedicated computing clusters.  The combined
use of supercomputers to generate large suites of gauge fields and to perform
the largest analysis jobs with these ensembles, and dedicated lattice capacity
hardware to perform the much larger volume of small- to medium-scale analysis
jobs, is the most cost-effective model for lattice-field theory calculations.
The successful utilization of future computing resources requires software
that runs efficiently on new computing architectures, and hence support for
postdocs and scientific staff to develop lattice gauge theory code.

Support of USQCD through hardware and software grants, access to
leadership-class computing facilities, and funding of lab and university
theorists, is essential to fully capitalize on the enormous investments in the
DOE's high-energy physics and nuclear-physics experimental programs.

Given continued support of the lattice gauge theory effort in the U.S. and
worldwide, lattice calculations will play a key role in definitively
establishing the presence of physics beyond the Standard Model and in
determining its underlying structure.

\section{Computing for perturbative QCD}
\label{chap:PQCD}
\subsection{Introduction}

One of the main challenges facing the particle physics community to date
is interpreting LHC measurements on the basis of accurate and
robust theoretical predictions.  The discovery of a Higgs-like
particle in summer 2012~\cite{Aad:2012tfa,Chatrchyan:2012ufa} serves
as a remarkable example of the level of detail and accuracy that must
be achieved in order to enable a
discovery~\cite{Dittmaier:2011ti,Dittmaier:2012vm,Heinemeyer:2013tqa}.
Signals for the Higgs boson of the Standard Model (SM) are orders of 
magnitude smaller than their backgrounds at the LHC, and they are 
determined by quantum effects.  Detailed calculations are therefore 
mandatory, and they will become even more necessary as we further 
explore the Terascale at the full LHC design energy.

Providing precise theoretical predictions has been a priority of the U.S.\ 
theoretical particle physics community for many years, and has seen an
unprecedented boost of activity during the last ten years. With the
aim of extracting evidence of new physics from the data, theorists
have focused on reducing the systematic uncertainty of their predictions
by including strong (QCD) and electroweak (EW) effects at higher orders
in the perturbative expansion. This is particularly important as
beyond-Standard-Model effects are expected at roughly the TeV scale. 
Typical decay chains of potential new particles would involve many 
decay products, several of which can be massive. The SM backgrounds 
are complex processes which call for highly sophisticated calculational 
tools in order to provide realistic predictions.

We have reached a time when no conceptual problems block us
from being able to break next-to-leading order (NLO)
perturbative QCD calculations into standard modular steps and automate
them, making them available to the worldwide LHC community.  It is
implicit that such an effort will benefit greatly from a unified
environment in which calculations can be performed and data can be
exchanged freely between theorists and experimentalists,
as well as from the availability of adequate computational means
for extensive multiple analyses.

We see the frontier of perturbative
calculations for collider phenomenology being in the
development and optimization of next-to-next-to-leading order (NNLO)
QCD calculations, sometimes combined with EW corrections, and in the
study of more exclusive signatures that requires resummation 
of logarithmically enhanced higher-order corrections to all orders.
It is also conceivable that techniques for matching NNLO fixed-order 
calculations to parton-shower simulations will be constructed in the 
next five years. In all cases, the availability 
of extensive computational resources could be instrumental
in boosting the exploration of new techniques as well as in
obtaining very accurate theoretical predictions at a pace and in a
format that is immediately useful to the experiments.

\subsection{Results and recommendations}

This planning exercise provided an incentive for implementing higher-order  
calculations in a standardized computing environment made available 
by DOE at 
NERSC.  Resource requirements were determined for the
calculation of important background and signal reactions at the
LHC, including higher order QCD and EW effects. Prototypical results 
are listed in Table~\ref{tab:summary} and have been summarized in a 
white paper~\cite{HPCWP}.

Different High Performance Computing (HPC) environments were tested
during this workshop and their suitability for perturbative QCD calculations 
was assessed. We find that it would be beneficial to make the national HPC 
facilities ALCF, OLCF, and NERSC accessible to particle theorists and
experimentalists so they can use existing
calculational tools for experimental studies involving extensive
multiple runs without depending on the computer power and manpower
available to the code authors. Access to these facilities will also
allow prototyping the next generation of parallel computer programs
for QCD phenomenology and precision calculations.

The computation of NLO corrections in perturbative QCD has been entirely
automated. Resource requirements for NLO calculations determined during 
this workshop can thus be seen as a baseline that enables phenomenology 
during the LHC era. NNLO calculations are still performed 
on a case-by-case basis, and their computing needs can only be 
projected with a large uncertainty. It seems clear, however, that cutting-edge 
calculations will require access to leadership class computing facilities.

The use of HPC in perturbative QCD applications is currently in
an exploratory phase. We expect that the demand for access to HPC
facilities will continue to grow as more researchers realize the 
potential of parallel computing in accelerating scientific progress. 
At the same time, we expect growing demand for educating young researchers 
in cutting-edge computing technology. It would be highly beneficial 
to provide a series of topical schools and workshops related 
to HPC in particle physics. They may be co-organized with experiments to foster 
the creation of a knowledge base.

Large-scale distributed computing in grid environments 
may become relevant for perturbative QCD applications 
in the near future. This development will be accelerated if computing grids
can also provide access to HPC facilities and clusters where parallel 
computing is possible on a smaller scale. The Open Science Grid (OSG)
has taken first steps in this direction, and we have successfully used their
existing interface. The amount of training for new users could be minimized
if the OSG were to act as a front-end to the national HPC facilities
as well as conventional computing facilities.

Additional details may be found in the full subgroup report \cite{Hoche:2013zja}.

\begin{table}
\begin{center}
  \begin{tabular}{ccc}
    \hline
    Type of calculation & CPU hours per project & Projects per year \\
    \hline\hline
    NLO parton level & 50,000 - 600,000 & 10-12\\
    NNLO parton level & 50,000 - 1,000,000 & 5-6\\
    Event generation & 50,000 - 250,000 & 5-8\\
    Matrix element method & $\sim$ 200,000 & 3-5\\
    Exclusive jet cross sections & $\sim$ 300,000 & 1-2\\
    Parton distributions & $\sim$ 50,000 & 5-6\\
    \hline
  \end{tabular}
\end{center}
  \caption{Summary of computing requirements for typical projects
    carried out by the U.S.~community~\cite{HPCWP}.
    \label{tab:summary}}
\end{table}

\section{Distributed computing and facility infrastructures}

Powerful distributed computing and robust facility infrastructures are essential for continued progress of particle physics across the Energy, Intensity, and Cosmic Frontiers.  Experiment and theory require a combination of HTC and HPC systems.  The LHC experiments are the dominant consumers of HTC.  They have been and will continue to be well served by it.  Most Intensity Frontier experiments can also be supported by HTC.  HPC is needed for applications such as lattice QCD, accelerator design and R\&D, data analysis and synthetic maps, N-body and hydro-cosmology simulations, supernova modeling, and, more recently, perturbative QCD.  Historically, national centers have focused primarily on HPC, but these centers have begun to address HTC, and are interested in attracting scientists who need HTC.

Energy Frontier experiments face a growth in data that will make it a challenge to meet their needs.  Doing so is possible, but it requires near-constant funding of the Worldwide LHC Computing Grid (WLCG), greater efficiencies in resource usage, and the evolution of software to take advantage of multicore processor architectures.   These experiments should also pursue and take advantage of opportunistic resources, be they in commercial clouds (which are not currently viable and cost effective as purchased resources), university and lab computing centers, or elsewhere.  The experiments would also benefit from further engagement with national HPC centers.  The centers could provide resources to particle physics experiments, and have support staff that could help port and integrate applications such as detector simulations that have not traditionally been used in HPC environments.

Intensity Frontier experiments have comparatively smaller computing needs.  There are no technical reasons why they could not be met.  Such experiments should  use resources available through the Open Science Grid (OSG) or at national computing centers.  They would benefit from a collective effort to gain access to resources and share software and training.

Cosmic Frontier experiments (and the simulations required to interpret them), lattice QCD, and accelerator design will need a large increase in HPC resources in the coming years. Demand for access to HPC across particle physics frontiers is expected to exceed the amount of available resources.  HPC-based computations are needed to interpret results from a number of important particle physics experiments, and to realize the scientific returns from the substantial investments in those experiments.  The NERSC report on particle physics computing needs \cite{NERSCHEP} indicates a shortage of HPC resources by a factor of four by 2017.  While funding and technology development needed to sustain traditional HPC growth rates are uncertain, they must be maintained to support particle physics.  There are a number of applications within particle physics that would benefit from exascale computing and a cadre of scientists eager to support efforts to reach that scale.

Distributed computing infrastructures, based at labs and universities, have been critical to the success of the Energy Frontier experiments and should continue to be able to serve these and other applications even as experiments grow.  There are no show-stoppers seen as scale increases, but various developments should be pursued to improve efficiency and ease of use.  Keeping sufficient staff support at a reasonable cost is a continuing concern; finding operational efficiencies could help address this.  Given that particle physics is the largest user of distributed scientific computing, currently in the form of HTC on computing grids, members of the field must continue to take a leadership role in its development.

National centers play an important role in some aspects of computing, and particle physics might be able to take advantage of an expanded role.  Experiments should explore the use of the HPC centers as part of their efforts to diversify their computing architectures.  These centers do have access to large, state-of-the-art resources, operational support, and expertise in many areas of computing.

We expect that distributed computing and facility infrastructures will continue to play a vital role in enabling discovery science.

Additional details may be found in the full subgroup report \cite{Bloom:2013yva}.

\section{Networking}

Particle physics research in all areas depends on the availability of reliable, high-bandwidth, feature-rich computer networks for interconnecting instruments and computing centers globally. Most particle physics-related data is transported by National Research and Education Networks (NRENs), supplemented by infrastructures dedicated to specific projects. NRENs differ from commercial networks, because they are optimized for transporting massive data flows generated by large-scale scientific collaborations. In addition, NRENs offer advanced capabilities --- such as multi-domain dedicated circuits --- which commercial providers do not have an incentive to deploy.

For decades, network traffic generated by particle physics has been a primary driver of NREN growth, and particle physics requirements have motivated NREN architectures and research activities. In the next ten years and beyond, the productivity of particle physics collaborations will continue to depend on an ecosystem of innovative global NRENs. 

Particle physics collaborations are now accustomed to viewing network transport as a reliable and predictable resource,  so much so that data models for ATLAS and CMS have evolved rapidly in response to NREN capabilities, but this state of affairs is not inevitable. Other data-intensive communities have begun to generate large traffic flows and, following the example of particle physics, to incorporate high-performance networks into science workflows. As a result of this broad trend toward data intensity across many disciplines, NRENs around the world will be challenged to meet the requirements of large-scale research, and must be adequately provisioned in order to continue serving the critical role they have played in the past.  

In support of our objectives through 2020, basic and applied networking research is necessary in a range of subjects. Critical questions include: 

\begin{itemize}
\item What future architectures will maximize utilization and minimize cost in core and campus networks?
\item How can emerging paradigms such as Software Defined Networking or Named Data Networking be harnessed most effectively to improve particle physics science outcomes?
\item Can networks evolve into adaptive, self-organizing, programmable systems that quickly respond to requests of particle physics science applications? 
\item If well-tuned host systems (or ensembles of them) have the ability to saturate a single backbone channel, what techniques and architectures can NRENs adopt to maximize data mobility?
\item How will the emerging “complexity challenge” arising from closer integration between networks and applications be managed, especially in the multi-domain, multi-national context?
\item How can diverse networks cooperate – automatically and securely – to offer science-optimized capabilities on a worldwide basis? 
\item Can discovery or automation techniques reduce the need for fragile, manual configuration? 
\item How will networks respond to the operational challenge of deploying  and managing dozens of wavelengths across large geographies under relatively flat funding prospects? 
\item Will post-TCP protocols become useful outside of highly-controlled demonstration projects? 
\item Would computer modeling of applications, networks, and data flows be useful in answering any of these questions?  
\item Will power consumption become a limiting economic or operational factor in this time period? 
\end{itemize}

Recent investments in network research have been insufficient.  Continued underfunding will compromise the ability of particle physics collaborations to maximize scientific productivity. Increased research funding, while necessary, is not sufficient; there also needs to be increased attention to the process of translating the results of network research into real-world architectures that NRENs can deploy and manage. Incentives and funding for such activities that enable deployment
are urgently needed. Because network research has now begun to intersect with research in services and applications, cross-disciplinary funding opportunities should also be available.  

A number of cultural and operational practices need to be overcome in order for NRENs (and global cyber-infrastructures more generally) to fully succeed. 
Expectations for network performance must be raised significantly, so that collaborations do not continue to design workflows around a historical impression of what is possible. 
The gap between peak and average transfer rates must be closed. 
Campuses must deploy secure science data enclaves – --- or  Science DMZs \cite{DMZ} – --- engineered for the needs of particle physics and other data-intensive disciplines.  Fortunately, each of these trends is currently underway, but momentum must be accelerated.   

Ten years from now, the key applications on which particle physics depends will only be fully successful, efficient, or cost-effective if they are run on the networks that exist today. 
During the next decade, research networks need to evolve into programmable instruments – --- flexible resources that can be customized for particular needs, but that exist within a common, integrated, ubiquitous framework that is reliable, robust and trusted for its privacy and integrity. These are major challenges, but they are tractable if funding agencies invest in innovative research, and maintain support for the exponential growth of NREN traffic. 

Additional details may be found in the full subgroup report \cite{Bell:2013fwa}.

\section{Software development, staffing, and training}

The success of particle physics will continue to critically depend on computing.
Managing the human activities associated with computing
(software development and management,
training and staffing) is an important part of that.  
Based upon our own experiences, and from
discussions with members of the particle physics community,
we have identified the following main goals for the next decade in the area
of software, staffing and training:

\begin{itemize}
\item Maximize the scientific productivity of our community
in an era of reduced resources, by using
software development strategies and staffing models that will result in products
that are useful for the entire particle physics community.
\item Respond to the evolving technology market, especially
with respect to computer processors, by
developing and evolving software that will perform with optimal efficiency
in future computing systems.
\item Insure that our developers and users will have the
training needed to create, maintain, and use the increasingly complex software
environments and computing systems that will be part of future 
particle physics projects.
\end{itemize}

Some specific recommendations we feel will help achieve these goals are detailed
below.

\begin{itemize}
    \item Software management, toolkits and reuse
    \begin{itemize}
        \item Continue to support established toolkits (such as Geant4, ROOT)
        \item Encourage the creation of new toolkits from existing successful common software (such as those for generators, tracking) 
        \item  Allow flexible funding of software experts to facilitate transfer of software and sharing of technical expertise between projects
        \item Facilitate code sharing through open-source licensing and use of publicly-readable repositories
        \item Consolidate and standardize software management tools to ease
migration of people from one project to another
    \end{itemize}

    \item Software development for new hardware architectures
    \begin{itemize}
        \item Invest in software needed to adapt to the evolution of computing processors, both as basic R\&D into appropriate
techniques and as re-engineering ``upgrades''
        \item Design new software and reengineer existing software to expose parallelism at multiple levels
        \item Develop flexible software architectures that can efficiently exploit a variety of possible future hardware options
    \end{itemize}    

    \item Staffing
      \begin{itemize}
          \item Recognize software efforts as sub-projects of the project
          \item Integrate computing professionals as part of the project team, over the life of the project or collaboration
          \item Integrate software professionals with scientist developers to insure software meets both the technical and scientific needs of the project
      \end{itemize}

    \item Training
    \begin{itemize}
        \item Use certification to document expertise and encourage the learning of
new skills
        \item Encourage training in software and computing as a continuing physics activity
        \item Use mentors to spread scientific software development standards
        \item Involve computing professionals in the training of scientific domain experts
        \item Use online media to share training
        \item Use workbooks and wikis as evolving, interactive software documentation
        \item Provide young scientists with opportunities to learn computing and software skills that are marketable for non-academic jobs
    \end{itemize}

\end{itemize}

Additional details may be found in the full subgroup report \cite{Brown:2013hwa}.

\section{Storage and data management}

The largest Energy Frontier experiments have developed, and are improving,
functional distributed data and workflow management systems that meet
their needs. These systems are expensive to develop and operate and
are thus  rarely appropriate for smaller experiments.

Particle physics currently benefits from, but can also be  constrained by, the
highly successful ROOT features supporting reading  and writing of
persistent data. No other major scientific field uses  ROOT or appears
interested in it. Major developments in the technology for
dealing with persistent data will
be required to take advantage of storage hardware on the  timescale of
LHC Run 3.

Particle physics should maintain and promote a vision of the future  in which fully
functional and low-operational-cost distributed computing and
persistency management is supported by software that is widely used in
data-intensive science.  To this end, developments in industry and the
wider  science community should be monitored actively. Particle physics should work
with the wider  science and computer science community to export and
adapt particle physics technologies and  vice-versa. In distributed computing, particle physics
should organize itself to significantly  reduce the number of diverse
approaches and share the ideas and  software developed
in the largest experiments with other activities where they are needed.

Rotating disk storage will suffer a marked slowdown in the  evolution
of capacity/cost.  This may be the largest perturbation of particle physics
computing  models that must attempt to optimize the roles of tape,
rotating disk, solid-state  storage, networking, and CPU.

Many of the components required to support virtual data  already exist
in the data and workflow management software of the largest
experiments.   The rigorous provenance recording required to support
the virtual data concept would  also benefit data preservation.

Computing model implementations should be flexible  enough to adapt to
a wide range of relative costs of the key elements of particle physics  computing.
In preparing for Run 3, the LHC program should seriously consider
virtual data as a way to accommodate scenarios where storage for
derived and  simulated data becomes relatively very costly.

All experiments across all frontiers
need  infrastructure that will allow scientists to store, catalog,
access, and  reprocess data sets years after the original physics
results are produced.  The inherent similarity of the requirements
across experiments and disciplines  calls for a coordinated investment
in common infrastructure to enable easy  access and adoption of best
practices in knowledge preservation.  Solutions  should be developed
that meet the needs of the particle physics and astrophysics  communities
before widespread release of data to the public can be expected  or
mandated.

Additional details may be found in the full subgroup report \cite{Butler:2013kka}.

\section{Conclusions}

For the {\bf Energy Frontier}, computing limitations already
reduce the amount of physics data that can be analyzed. The
planned upgrades to the LHC energy and luminosity are expected to result in
a ten-fold increase in the number of events and a ten-fold increase in
event complexity. Efforts have begun to increase code efficiency and
parallelism in reconstruction software and to explore the potential of
computational accelerators such as GPUs and Xeon Phi.
Saving more raw events to tape and only
reconstructing them selectively is under consideration. 
The LHC produces about 15 petabytes (PB) of raw data per
year now, but in 2021 the rate may rise to 130 PB. Attention needs to be
paid to data management and wide-area networking, to assure that network
connectivity does not become a bottleneck for distributed event
analysis. It is important to monitor storage cost and throughputs. More
than half of the computing cost is now for storage, and in the future it
may become cost-effective to recalculate certain derived quantities rather
than storing them.

{\bf Intensity Frontier} experiments have combined computing requirements
on the scale of a single Energy Frontier experiment, but they are a more
diverse set than those of the Energy Frontier.  
We conducted a survey and found that
there is significant commonality in different experiments' needs. Sharing
resources across experiments, as in the Open Science Grid, is a first step
in addressing peak computing needs.  Continued coordination of
software development among these experiments will increase efficiency of
the development effort.
Leveraging the data handling experience and
expertise of the Energy Frontier experiments for the diverse Intensity
Frontier experiments would significantly improve their ability to reconstruct
and analyze data.

{\bf Cosmic Frontier} experiments will greatly expand their data volumes needs
with the start of new surveys and the development of new instruments.
Current data sets are about 1 PB, and the total data set is expected to be
about 50 PB in ten years. Beyond that, in 10--20 years data will be
collected at the rate of 400 PB/yr. On the astrophysics and cosmology
theory side, some of the most challenging simulations are being run on
supercomputers. 
Current allocations for this effort are approximately 200M core-hours annually.
Very large simulations will require increasing computing
power. Comparing simulations with observations will play a crucial role in
interpretation of experiments, and simulations are needed to help design
new instruments. There are very significant challenges in dealing with new
computers' architectures and very large data sets, as described above.
Growing archival storage, visualization of simulations, and allowing public
access to data are also issues that need attention.

{\bf Accelerator science} is called on to simulate new accelerator designs
and to provide near-real-time simulations feedback for accelerator
operation. 
Research into new algorithms and designs has the potential to bring new ideas and
capabilities to the field.
It will be necessary to include additional physics in codes and
to improve algorithms to achieve these goals. Production runs can use from
10K to 100K cores. Considerable effort is being expended to port to new
architectures, in particular to address the real-time requirements.

{\bf Lattice field theory} calculations rely on national supercomputer
centers and hardware purchased for the USQCD Computing Project. Allocations
at supercomputer centers have exceeded 500 M core-hrs this year, and
resource requests will go up by a factor of 50 by the end of this decade.
This program provides essential input for interpretation of a number of
experiments, and increased precision will be required in the future. For
example, the $b$ quark mass and the strong coupling $\alpha_s$ will need to
be known at the 0.25\% level, a factor of two better than now, to compare
precision Higgs measurements at future colliders with 
Standard Model predictions.  Advances
in the calculation of hadronic contributions to muon $g-2$ will be needed
for interpretation of the planned experimental measurement.

{\bf Perturbative QCD} is essential for theoretical understanding of
collider physics rates. Codes were ported to the HPC centers at NERSC and
OLCF, and also run on the Open Science Grid. They have also been
benchmarking GPU codes and finding impressive speed up with respect to
 a single core.
A computer at CERN was used to benchmark the Intel Xeon Phi chip.
A repository of codes has been established at NERSC.  A long term goal is
to make it easy for experimentalists to use these codes to compute Standard
Model rates for the processes they need.

The {\bf Distributed computing and facilities infrastructures} subgroup
looked at the growth trends in distributed resources as provided by the
Open Science Grid, and the national high performance computing (HPC)
centers. Most of the computing by experiments is of the HTC type, but HPC
centers could be used for specific work flows. Using existing computing
centers could save smaller experiments from large investments in hardware
and personnel. Distributed HTC has become important in a number of science
areas outside particle physics, but particle physics is still the 
biggest user and must continue to
drive the future computing development. HPC computing needs for theoretical
physics will require an order of magnitude increase in capacity and
capability at the HPC centers in the next five years, and two orders of
magnitude in the next ten years.

The {\bf Networking} subgroup considered the implications of distributed
computing on network needs, required R\&D and engagement with the National
Research and Education Networks (which carries most of our traffic). A
number of research questions were formulated that need to be answered
before 2020. Expectations of network performance should be raised so that
planning for network needs is on par with that for computing and storage.
The gap between peak bandwidth and delivered bandwidth should be narrowed.
It was not felt that wide-area network performance will be an
insurmountable bottleneck in the next five to ten years as long as
investments in higher performance links continue. However, there is
uncertainty as to whether network costs will drop at the same rate as they
have done in the past.

The {\bf Software development, personnel, and training} subgroup has a
number of recommendations to implement three main goals. The first goal is
to use software development strategies and staffing models that result in
software more widely useful to the particle physics community. The second goal is to
develop and support software that will run with optimal efficiency on
future computer architectures. The third goal is to insure that developers
and users have the training necessary to deal with the increasingly complex
software environments and computing systems that will be used in the future.

The {\bf Storage and data management} subgroup found that storage continues
to be a cost driver for many experiments. It is necessary to manage the
cost to optimize the science output from the experiment. Tape storage
continues to be relatively inexpensive and should be more utilized within
the storage hierarchy. 
Disk storage is likely to increase in capacity/cost relatively slowly due
to a shrinking consumer market and technology barriers.
It  can be costly for experiments to operate their own
distributed data management systems, thus
continued R\&D in this area would benefit a number of experiments.

To summarize, the challenging resource needs for the planned and proposed
physics programs require efficient and flexible use of all resources. 
Particle physics
needs both distributed HTC and HPC.  Emerging experimental programs might
consider a mix to fulfill demands. 
Programs to fund these resources need to continue. 
It may also be possible to use shared computer resources and opportunistic 
sources of computing to meet some needs.  Commercial cloud providers
may also provide a useful resource, particularly if prices are reduced.
There is increasing need for data-intensive computing in traditionally
computation-intensive fields, including at HPC centers.  

In order to satisfy our increasing computational demands, the field needs
to make better use of advanced computing architectures. With the need for
more parallelization, the complexity of software and systems continues to
increase, impacting architectures for application frameworks, workload
management systems, and also the physics code. We must develop and maintain
expertise across the field, and re-engineer frameworks, libraries, and
physics codes. Unless corrective action is taken to enable us to take full
advantage of the new hardware architectures, we could be frozen out of 
cost-effective computing solutions on a time scale of 10 years. There is a large
code base that needs to be re-engineered, and we currently do not have
enough people trained to do it.

The continuing huge growth in observational and simulation data drives the
need for continued R\&D investment in data management, data access methods,
and networking. Continued evolution of the data management and storage
systems will be needed in order to take advantage of new network
capabilities, ensure efficiency and robustness of the global data
federations, and contain the level of effort needed for operations.
Significant challenges with data management and access remain, and research
into these areas could continue to bring benefit across the Frontiers.  

Network reliability is essential for data intensive distributed computing.
Emerging network capabilities and data access technologies improve our
ability to use resources independent of location. This will enable use of
diverse computing resources including dedicated facilities, university computing 
centers, resources shared opportunistically between PIs, and potentially 
also commercial clouds. Leadership-class HPC centers may also become 
relevant for data-intensive computing.
The computing models should
treat networks as a resource that needs to be managed and planned for.

Computing will be essential for progress in theory and experiment over the
next two decades. 
The advances in computer hardware that we have seen
in the past may not continue at the same rate in the future. The issues
identified in this report will require continuing attention from both the
scientists who develop code and determine what resources best meet
their needs, and from the funding agencies who will review plans and
determine what shall be funded.
Careful attention to the computational challenges in our field
will increase efficiency and enable us to meet the
experimental and theoretical physics goals identified through the Snowmass process.

\end{document}